\documentclass{article}

\usepackage[final,nonatbib]{neurips_2021_ml4ps}

\usepackage[utf8]{inputenc} 
\usepackage[T1]{fontenc}    
\usepackage{url}            
\usepackage{booktabs}       
\usepackage{amsfonts}       
\usepackage{nicefrac}       
\usepackage{microtype}      
\usepackage{hyperref}
\usepackage{xcolor}         
\usepackage{graphicx}
\usepackage{tikz}
\usetikzlibrary{quantikz}
\usepackage{caption}
\usepackage{subcaption}
\newcommand{\comment}[1]{}
\title{Quantum Machine Learning for Radio Astronomy}

\author{%
  Mohammad Kordzanganeh \\
  Department of Physics \& Astronomy\\
  University of Manchester, UK\\
  \texttt{mohammad.kordzanganeh@gmail.com} \\
  \And
  Aydin Utting \\
  Department of Physics \& Astronomy\\
  University of Manchester, UK\\
  \texttt{aydinutting@gmail.com } \\
  \AND
  Anna Scaife\thanks{The Alan Turing Institute, 96 Euston Rd, London, UK \texttt{a.scaife@turing.ac.uk}} \\
  Department of Physics \& Astronomy\\
  University of Manchester, UK\\
  \texttt{anna.scaife@manchester.ac.uk} \\
}

\begin{document}

\maketitle

\begin{abstract}
  In this work we introduce a novel approach to the pulsar classification problem in time-domain radio astronomy using a Born machine, often referred to as a \emph{quantum neural network}. Using a single-qubit architecture, we show that the pulsar classification problem maps well to the Bloch sphere and that comparable accuracies to more classical machine learning approaches are achievable. We introduce a novel single-qubit encoding for the pulsar data used in this work and show that this performs comparably to a multi-qubit QAOA encoding.
\end{abstract}

\section{Introduction}
\label{sec:intro}

Pulsars are rapidly rotating neutron stars that emit very precisely timed repeating radio pulses with periods of milli-seconds to seconds. These objects are formed through the death of massive stars ($>8$\,M$_{\odot}$), which have collapsed masses that are insufficient to undergo complete gravitational collapse and form a black hole but are sufficiently massive that collapse causes their electrons to combine with protons and form neutrons, a process which continues until neutron degeneracy pressure is high enough to prevent further gravitational collapse. By the time this happens such stars are almost exclusively comprised of neutrons, compressed into a sphere of approximately 20\,km in diameter \cite{pulsarbook}. 

Finding and observing pulsars is a core science aim of the Square Kilometre Array (SKA) observatory, a global project to build the world's largest radio telescope \cite{skapulsar}. The SKA intends to conduct a cosmic census of the pulsar population with the aim of addressing a number of key questions in modern physics, including the detection and mapping of gravitational waves as they pass through our Galaxy. By timing the arrival of radio pulses produced by numerous pulsars with milli-second spin periods, the presence of gravitational waves can be detected as a disturbance in the regularity of these pulse arrival times on Earth \cite{foster1990,hobbs2010,janssen2015}, opening up a new gravitational wave regime to that detected by (e.g.) the LIGO experiment on Earth \cite{ligo}. To conduct such an experiment the SKA must identify and map the location of thousands of previously unknown pulsars by separating their periodic signals from those of other radio frequency interference sources. Consequently the development of classification algorithms for this purpose has become a subject of significant interest over the past few years \cite{lyon2016}. 

\vskip .1in
\noindent
\textbf{This work} In this work we introduce a novel approach to pulsar classification using a Born machine, often referred to as a \emph{quantum neural network}. Using a single qubit architecture, we show that the pulsar classification problem maps well to the Bloch sphere and that comparable accuracies to more classical machine learning approaches are achievable. We introduce a novel single-qubit quantum encoding for the pulsar data used in this work and compare this encoding with the more standard QAOA encoding \cite{qaoa_origin_paper}. The former shows great promise in trainability and expressivity compared to the QAOA ansatz. 

\section{Quantum Model}\label{sec:quantum_model}
\begin{figure}
    \centering
    \begin{quantikz}
\lstick{\ket{0}} & \gate{H}
& \gate{W^{(0)}} & \gate{R_z(x_1)} & \gate{W^{(1)}} & \qw \cdots \qw& \gate{W^{(n-1)}}
& \gate{R_z(x_n)} & \gate{W^{(n)}}
& \meter{}
\end{quantikz}
    \caption{One encoding repetition of the single qubit network used in this work. To add more repetitions the series of gates (excluding the initial Hadamard gate) need to be repeated with new trainable parameters.}
    \label{fig:quantum_model_encoding}
\end{figure}
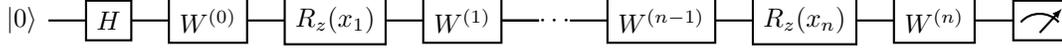

The quantum model introduced in this work is an extension of the single-qubit models explored in \cite{schuld_fourier}. These extensions are: (i) a model for a general single-qubit trainable layer, and (ii) the extension of the single feature Fourier series to a multi-dimensional Fourier series. The architecture of the model consists of generalised trainable layers interlaced with a single-gate rotations encoding the features, see Figure~\ref{fig:quantum_model_encoding}.

\vskip .1in
\noindent
\textbf{Trainable layers} The trainable layers play an important role in determining the expressivity of a quantum model \cite{schuld_fourier}. A trainable layer needs to be complex enough to provide sufficient access to the Fourier coefficient space, but also simple enough to be efficient for gradient calculations. The trainable layer chosen in this work consists of three consecutive rotations, about $Z$, then $X$, and then $Y$. This gives the most general $SU(2)$ rotation, and in a single-qubit case it can convert any initial state, $|\psi\rangle$, to any final state, $|\phi\rangle$. This ensures that we can fully access the group space and maximally access the Fourier space in the most efficient way possible\footnote{Given a particular use case, these trainable layers may be pruned to increase efficiency.}.

\vskip .1in
\noindent
\textbf{Multi-feature encoding} The data features are encoded using $Z$-rotations placed between the trainable layers.  This means that for the 8 features in the HTRU\,2 dataset, there are 9 trainable layers enclosing 8 $Z$-rotation encoding gates.  We show here that this architecture is able to express the first degree multi-dimensional Fourier series of the dataset. We refer to this as a quantum asymptotically universal multi-feature (QAUM) encoding.

We follow the analytical derivation of \cite{schuld_fourier} by encoding using
\begin{equation}
S(x)=e^{i\mathcal{G}x},
\end{equation}
where $\mathcal{G}$ is an $SU(2)$ generator, i.e. one of the Pauli matrices. These encoding layers are surrounded by trainable layers with weights, $W$.  In tensor notation, we can express a circuit with a single encoding:
\begin{equation}\label{eqn:single_enc_tensor}
|\psi\rangle = W^{(2)}S(x)W^{(1)}|0\rangle
\longrightarrow W^{(2)}_{ki}e^{i\mathcal{G}_{ij}x}W^{(1)}_{j1},
\end{equation}
where $|0\rangle \longrightarrow [1,0]^T$ is absorbed in the second index of $W^{(1)}$ and the Einstein summation convention is assumed. Choosing $\mathcal{G}$ to be the Pauli-Z matrix, we can re-write Eqn \ref{eqn:single_enc_tensor} with the eigenvalues of Pauli-Z $\lambda \in \{-1,1\}$:
\begin{equation}
|\psi\rangle_k \longrightarrow W^{(2)}_{ki}e^{i\lambda_{i}x}W^{(1)}_{i1} = \left(W^{(2)}_{ki}W^{(1)}_{i1}\right)e^{i\lambda_{i}x}.
\end{equation}
This encoding can be extended to a more general case with $N$ features and $L$ encoding repetitions:

\begin{equation}
|\psi\rangle_k \rightarrow W^{(L^1)}_{ki_{L^1}} e^{i \lambda_{i_{L^1}}x_1} W^{(L^2)}_{i_{L^1}i_{L^2}} 
\cdots W^{(L^N)}_{i_{L^{N-1}}i_{L^N}} e^{i \lambda_{L^N}x_N} W^{((L-1)^N)}_{i_{L^{N}}i_{(L-1)^1}} 
\cdots W^{(0)}_{i_{1^{N}}1}.
\end{equation}
Measuring this qubit in some basis defined by a measurement operator, $\hat{M}$, gives us the expectation value of this operator in the $|\psi\rangle$ basis\footnote{In this work, $\hat{M}$ is taken to be the fundamental measurement basis of $\{|0\rangle,|1\rangle\}$},
\begin{equation}
\label{eqn:expectation_value}
\langle\psi| \hat{M} | \psi \rangle 
=
W^{\dagger(0)}_{1j_{1^{N}}} \cdots
W^{\dagger(L^1)}_{j_{L^1}k'}
M_{k'k}
W^{(L^1)}_{ki_{L^1}}
 \cdots W^{(0)}_{i_{1^{N}}1}
  e^{i\sum_{l=1}^{N} x_l \left(\sum_{m=1}^{L} \lambda_{i_{m^l}} -  \sum_{p=1}^{L}\lambda_{j_{p^l}}\right)}.
\end{equation}
In this equation, it is possible to think of $\lambda_{i_{m^l}}$ as a matrix whose dimensions are indexed by $m$ and $l$. The values in this matrix are $1$ and $-1$ with uniform probability and all the possible combinations of these values, create a matrix ensemble. Each member of the ensemble has an associated coefficient that is calculated by multiplying the trainable layers with indices $\textbf{i}_{m^l}$ - a vector of $1$s and $2$s. All of these possibilities are included in the calculation of the expectation value in Equation~\ref{eqn:expectation_value}. With this in mind, we can re-write the exponent of this equation
\begin{equation}
\beta = \sum_{l=1}^{N} x_l \left(\sum_{m=1}^{L} \lambda_{i_{m^l}} -  \sum_{p=1}^{L}\lambda_{j_{p^l}}\right)
 = 
 [1,1,\cdots,1] ([\lambda_{i_{m^l}}]-[\lambda_{j_{p^l}}]) [x_1,x_2,\cdots,x_N]^T,
\end{equation}
where the sum over $m$ and $p$ are replaced with a matrix multiplication of the member ensembles by a vector of ones, and the sum over $l$ by a multiplication with the vector of features. Making the substitution $2 \alpha = \lambda_{i_{m^l}} - \lambda_{i_{p^l}}$ shows that members of $\alpha$ are $\{-1,0,1\}$, where the number of terms contributing to each value is distributed as $(1,2,1)$. Multiplication of $\alpha$ with the vector of ones yields 
\begin{equation}
\gamma = [1,1,\cdots,1] \alpha = [\gamma_1,\gamma_2,\cdots,\gamma_N] ,
\end{equation}
where $\gamma_q \in \{-L,-(L-1),\cdots,-1,0,1,\cdots,L-1,L\}$, and subsequently the number of terms contributing to each is given by $\{{\binom{2L}{2L}},{\binom{2L}{2L-1}},{\binom{2L}{2L-2}},\cdots,{\binom{2L}{0}}\}$. This means that the higher frequency terms are less accessible than those with lower frequencies, and consequently
\begin{equation}
\beta = 2 \gamma \left[x_1,x_2,\cdots,x_N\right]^T.
\label{eqn:2beta}
\end{equation}
We get a superposition of all the frequencies of the features up to $L$, the number of encoding repetitions. This means that by increasing the number of repetitions, we approximate the problem with a truncated Fourier series of higher frequency. We therefore expect the training loss to decrease as the number of repetitions increase.   

\subsection{Training}
\label{sec:training}

For this work we use the HTRU\,2 pulsar dataset\footnote{Data are publicly available at \url{https://archive.ics.uci.edu/ml/datasets/HTRU2}.} \cite{lyon2016}, which contains 16,259 spurious examples caused by RFI/noise and 1,639 real pulsar examples. Each example is described by 8 continuous feature variables and a single class label. We scale the original HTRU\,2 feature values individually to lie in the range $(0,\pi)$ to produce the full Fourier series\footnote{The reason $\pi$ was chosen over $2\pi$ was to account for the factor of $2$ in Eqn \ref{eqn:2beta}.}.

The circuits were simulated using the PennyLane \textit{default.qubit} \cite{pennylane_doc} device and the gradients were calculated using the simulator's backprop method. The Adam Optimiser with a learning rate of 0.1 was utilised to update the trainable parameters and the cross entropy loss function was used. The full dataset was randomly sampled 5 times, to obtain five training data sets with 100 data points each and a balanced class ratio. The model was trained for 150 epochs on Kaggle using Intel Xeon CPU. Each training run took approximately $\Delta t = 760\pm20$ seconds\footnote{Time shown is for the QAUM $L=2$ model, which is the main subject of comparison in Section \ref{sec:results}.  The training time is highly dependent on the number of trainable parameters.}\footnote{Full code to recreate the results is available at \url{https://github.com/kordham/qaum}.}.

Two types of uncertainty were evaluated: (i) weight initialisation error, and (ii) sampling error.  The former was measured by running the training on a set sample of data 5 times using uniformly randomised weights in the range $w=[0,2\pi]$, and the latter was measured by changing the training sample 5 times. In each case, the mean and standard deviation of the performance was calculated. 

The QAOA ansatz using $Y$ rotations was directly imported from the PennyLane templates for training the network.  This model was implemented on 9 qubits and had 18 trainable parameters per repetition.

\section{Results}
\label{sec:results}

\begin{figure}
     \centering
     \begin{subfigure}[b]{0.22\textwidth}
         \centering
         \includegraphics[width=\textwidth]{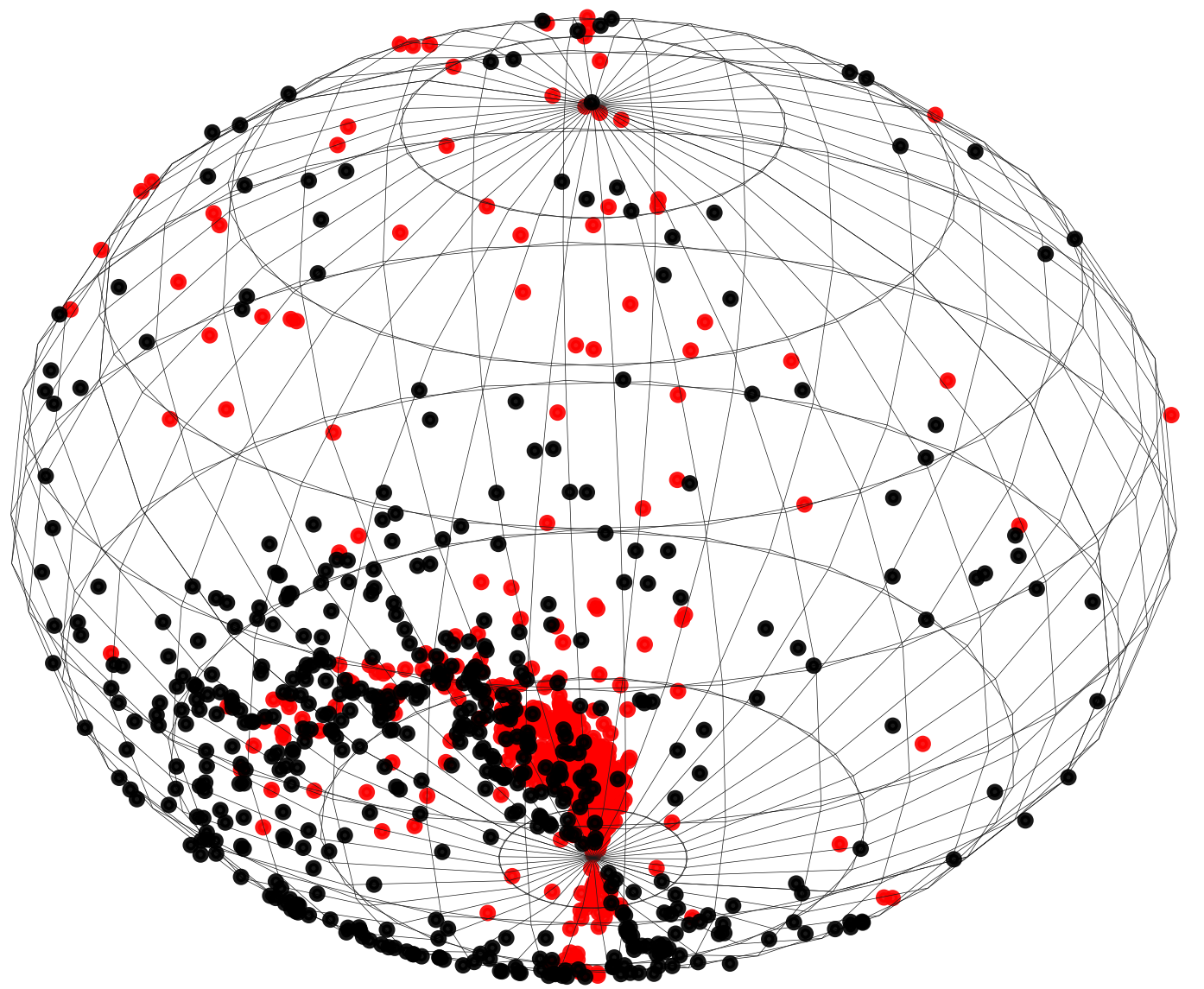}
     \end{subfigure}
     \hfill
     \begin{subfigure}[b]{0.22\textwidth}
         \centering
         \includegraphics[width=\textwidth]{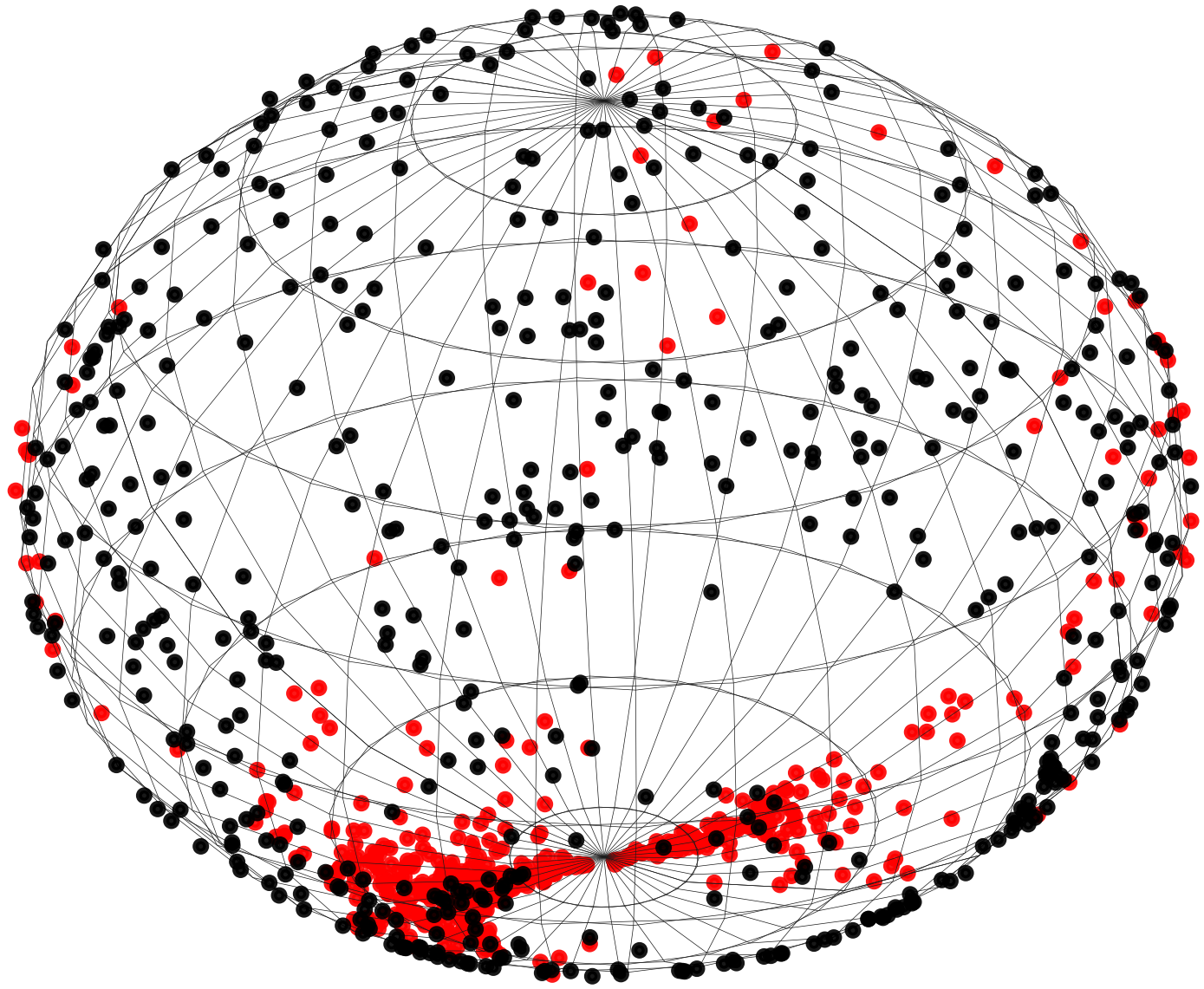}
     \end{subfigure}
     \hfill
     \begin{subfigure}[b]{0.22\textwidth}
         \centering
         \includegraphics[width=\textwidth]{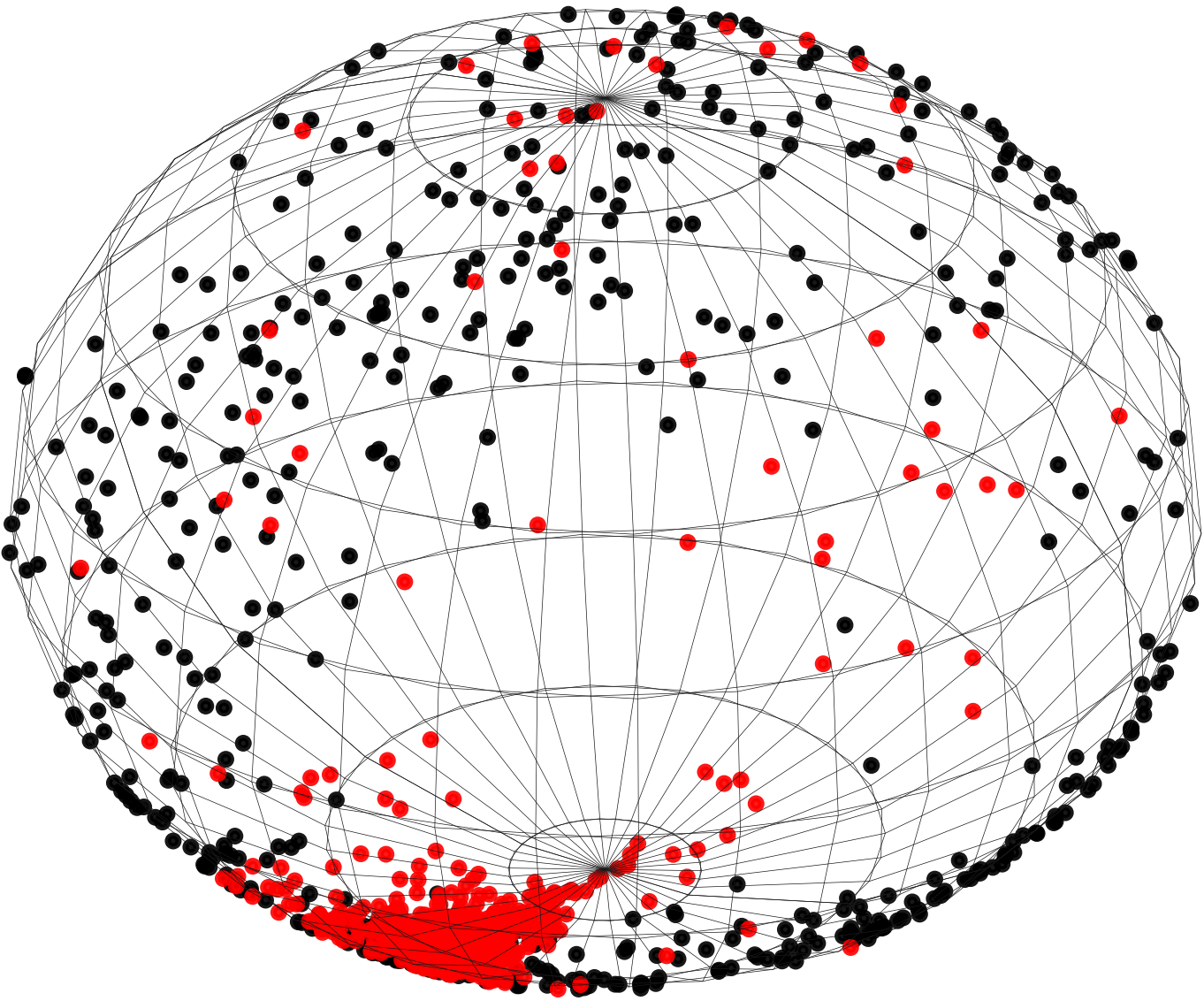}
     \end{subfigure}
     \hfill
     \begin{subfigure}[b]{0.22\textwidth}
         \centering
         \includegraphics[width=\textwidth]{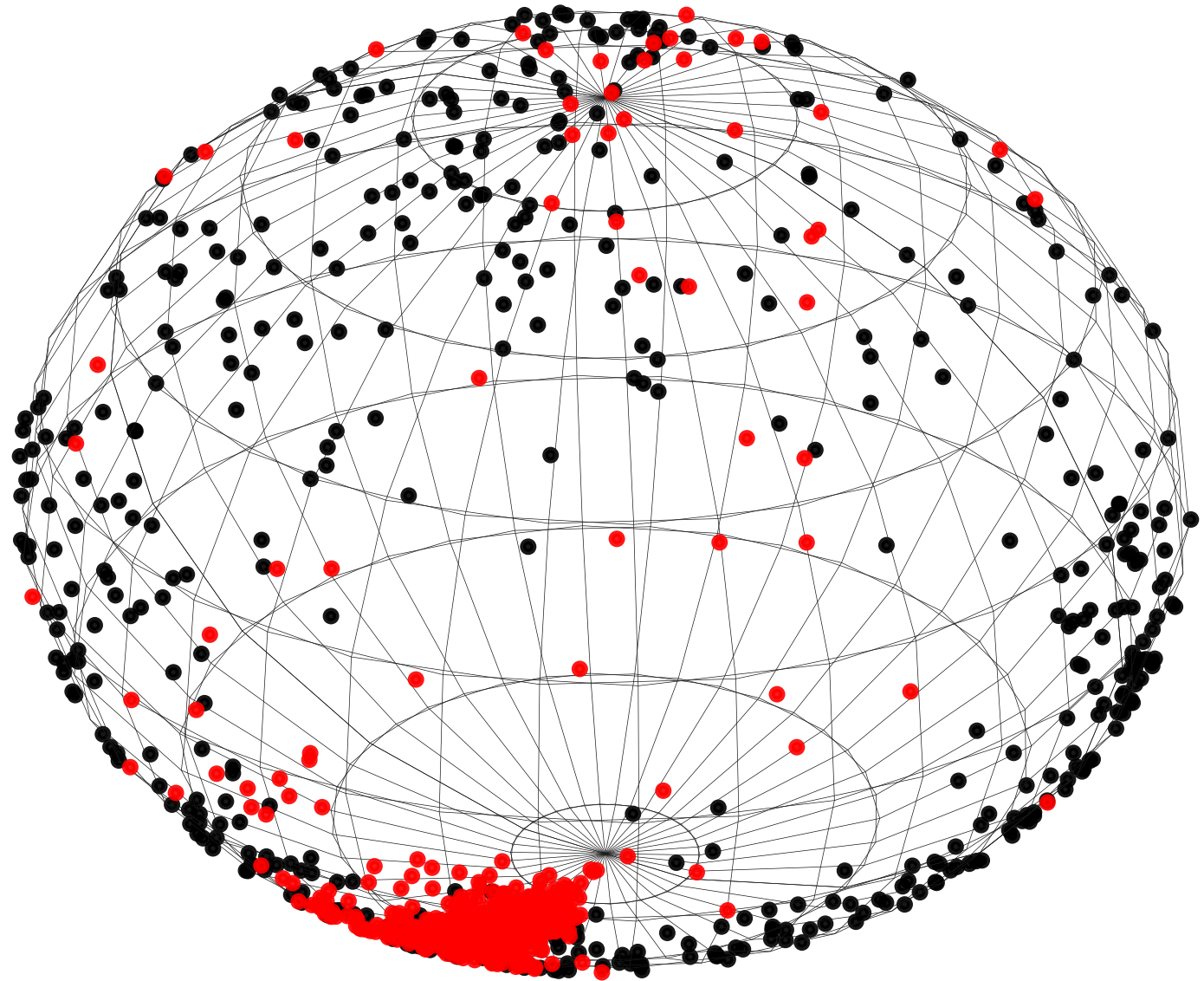}
     \end{subfigure}
        \caption{The mapping of the HTRU\,2 dataset to the Bloch sphere using the QAUM $L=2$. The spheres from left to right show the status of mapping for epochs 1, 50, 100 and 150, respectively. The red data points show the pulsars, and the black ones show the non-pulsar training data. To produce these spheres 1000 data points were used to showcase the mapping more clearly. It is evident that training the model makes it gather the pulsar data closer to the south pole of the sphere (the $|1\rangle$ state).}
        \label{fig:bloch_sphere}
\end{figure}

The trained QAUM $L=2$ model achieves a training accuracy of $95.8\pm2.3\%$ and a test accuracy of $91.6\pm3.6\%$, approaching the accuracies achieved using more classical machine learning approaches on the full HTRU\,2 dataset \cite{lyon2016}. Figure~\ref{fig:bloch_sphere} shows how the QAUM encoding maps the HTRU\,2 dataset to the Bloch sphere during training. QAOA $L=3$ achieves a training accuracy of  $91.2\pm2.3\%$, and a similar test accuracy of $91.6\pm1.9\%$.

We note that although the QAUM ansatz may look like a recurrent structure, the ordering of the features in QAUM is arbitrary. To confirm this, we repeated the experiment with randomly shuffled features and recovered the same performance.  This can also be demonstrated theoretically due to the symmetry of the features: in Equation~8 the order of the features could be permuted and the $\gamma$ matrix would still distribute the term contributions in the same manner.

Table~\ref{tab:performance_res} shows the results of the QAOA ansatz compared with this work for encoding repetitions of $L=\{1,2,3\}$. The number of parameters for each model increases differently, but they are closest in $L=3$ repetitions of QAOA and $L=2$ repetitions of QAUM (this work).  Comparing these, the latter achieves a significantly lower minimum loss with fewer trainable parameters and qubits. We speculate that this difference in performance may be due to missing Fourier frequencies in the QAOA ansatz. Furthermore, the Fourier frequencies that do exist may not fully accessed because they do not explore the entire group space for a 9-qubit QAOA.

These results could mean that there is an advantage in using a single qubit for datasets with any number of features without compromising on model performance.   On the other hand, the single qubit model is a deeper quantum circuit\footnote{The depth of a quantum circuit is defined as the number of quantum gates applied to its most manipulated qubit. The QAUM ansatz is often deeper as it encodes all the features in a single qubit.}.  This means that the desirability of this model depends on the fidelity of the device and its coherence time.  For devices with fewer qubits but longer coherence times such as the recent Honeywell's System Model H1 experiment \cite{honeywell-trapped-ion} this could be an appropriate model to use.

\begin{table}[]
\begin{tabular}{ccccccc}

\textbf{Model} & \multicolumn{1}{l}{\textbf{Qubits}} & \multicolumn{1}{l}{\textbf{Repetitions}} & \multicolumn{1}{l}{\textbf{Params}} & \multicolumn{1}{l}{\textbf{Min Loss}} & \multicolumn{1}{l}{\textbf{Initialisation Err}} & \multicolumn{1}{l}{\textbf{Sampling Err}} \\ \hline
QAOA          & 9                                   & 1                                        & 18                                  & 0.603                                 & 0.001                                           & 0.052                                     \\
\textbf{}     &                                     & 2                                        & 36                                  & 0.365                                 & 0.004                                           & 0.027                                     \\
\textbf{}     &                                     & \textbf{3}                               & \textbf{54}                         & \textbf{0.346}                        & \textbf{0.003}                                  & \textbf{0.056}                            \\ 
QAUM     & 1                                   & 1                                        & 27                                  & 0.276                                 & 0.010                                           & 0.030                                     \\
(this work)     &                                     & \textbf{2}                               & \textbf{51}                         & \textbf{0.251}                        & \textbf{0.018}                                  & \textbf{0.038}                            \\
\textbf{}     &                                     & 3                                        & 75                                  & 0.208                                 & 0.012                                           & 0.045                                     \\ \hline
\end{tabular}
\caption{The minimum training losses and their uncertainties after $150$ epochs of training for the QAOA model and the QAUM model (this work). It is notable that the initialisation error is higher for this work than the QAOA ansatz by a considerable margin.}
\label{tab:performance_res}
\end{table}

\section{Conclusions}
\label{sec:conc}

This work has looked at a novel approach to classifying the HTRU\,2 pulsar dataset using quantum machine learning.  We extend the methods used previously in \cite{schuld_fourier} to create a single qubit quantum model that may be applied to any general binary classification problem, regardless of the number of features.  This network was compared with an established quantum variational solving circuit known as the QAOA ansatz on this dataset.  The single qubit network trained to a lower loss than the QAOA despite the large difference in the number of qubits.

We show that the single-qubit encoding creates a multi-dimensional Fourier series whose highest frequency is determined by the number of repetitions.  To access the maximum potential of the Fourier coefficients, this work suggests the use of the most general state of a qubit as the trainable layers. 

We note that although the pulsar classification application considered here is not a high-dimensional problem, this does not necessarily mean that this architecture is limited by the Bloch sphere. Indeed, by adding additional repetitions we generate more Fourier terms which should assist in separating classes in a given classification task. The limitation of this, however, is the accessibility to the Fourier space. The performance of a  single-qubit QAUM, as demonstrated here, compared with a 2-qubit QAUM is not immediately clear, and is the subject of future research. Furthermore, while the single qubit encoding demonstrated here can be efficiently run on a classical computer - there are currently few arguments that make it appealing to run on a quantum computer - this would no longer be the case when extending the QAUM ansatz to the multi-qubit case.

\begin{ack}
The authors gratefully acknowledge support from the UK Alan Turing Institute under grant reference EP/V030302/1.
\end{ack}


\bibliography{main_arxiv.bib}
\bibliographystyle{abbrv}

\clearpage
\end{document}